# OPTICAL PROPERTIES OF NANOMETER-SCALE STRUCTURES

T.A.Kudykina, A.I.Pervak

University "Ukraina", 1-G Horiva Street, Kyiv 04071, Ukraine
e-mail: tkudykina@ukr.net; pervak@gmail.com

**ABSTRACT**

Two approaches (micro- and macro- investigations) are used to determine the dimension dependences of the optical parameters of the nanometer-scale layers of materials. It is shown that both an index of refraction and coefficient of absorption depend strongly on the thickness of the layer. In this region of thicknesses, the dimension resonance occurs, where an index of refraction has a maximum and a coefficient of absorption has a minimum.

The numerical calculation of the optical parameters of some materials (*Ag, Al, Fe, Ge, Si, Se, Te*) have been carried out with the use of the experimental data of reflection and transparency of thin layers, obtained in a series of works, and with our formulas for the wave amplitudes and the laws of refractions. The analogues of the Fresnel formulas and the Snell law have been derived from the Maxwell boundary conditions where the absorption and conductivity of media were taken into account. The use of our formulas for the wave amplitudes leads to the fulfillment of the conservation law of energy both for the TE- and TM- polarizations of light.

The reemission of light is possible for the thicknesses, which is equal to the wavelength of light in a medium, divisible by integer: $\frac{2\pi d_{res}}{2} = \lambda_m \cdot m = \frac{\lambda_0}{n} \cdot m$, where $m = 1,2,3\ldots$ The visible luminescence of the porous silicon, silicon whiskers and other nanometer –scale structures can be explained by their emission in the case of dimension resonance because in this region of thicknesses, layers have a big negative coefficient of absorption and little – of a damping.

It is also shown that in the case of the TM-polarization, in a material with a high conductivity, light refracts in a "wrong" direction (at all positive optical parameters) due to a surface current. In the case of the TE-polarization, light refracts only in a "right" direction.

## 1. INTRODUCTION

Investigation of thin layers of semiconductors, metals and dielectrics, which have been carried out in the middle of the twentieth century, has shown that their optical properties depend strongly upon their thickness $d$ and only at $d > \frac{\lambda_0}{n}$ (where $\lambda_0$ – is a wavelength of light in vacuum, and $n$ – is an index of refraction) they aspire to their volume meanings [1,2 and cited works].

The most remarkably this difference is seen in the luminescence. Usually, the visible luminescence is observed from very thin films and nanometer-scale structures of other shapes both from the metals and semiconductors. The visible luminescence was also observed from Yakut diamonds: pure diamonds did not have any luminescence, the crystals with a few amount of impurities had blue and green luminescence, and the crystals, which had a lot of impurities (they were seen by the naked eye), had red or orange luminescence.

Several decades ago, it was noted that a thin nanometer-scale silver film ($d \sim 5\text{-}6$ nm) on a glass substrate absorbs and irradiates effectively the visible light [3] and the same free film does not irradiate [4]. Perhaps, it is due to the fact, that emission is determined by Poynting's vector

$$\vec{S} = \frac{c}{4\pi}[\vec{E} \times \vec{H}]$$

and an electric field is close to zero in metals.

About sixteen years ago, Canham [5] as well as Lehmann and Gösele [6] independently reported the remarkable effect of strong photoluminescence from porous silicon (PS) which was produced by anodic oxidation (in HF/ $C_2H_5OH$ electrolyte) of crystalline silicon. This fact was considered incompatible with the indirect gap of a bulk crystalline silicon at $E_g = 1.12$ eV at room temperature (near infrared).

Visible luminescence was also observed from nanometer-scale silicon (or germanium) crystallite, nanofabricated silicon-germanium substrate, porous silicon obtained by anodization or without anodization, and silicon substrate irradiated by an ion beam with a subsequent HF-solution treatment, from silicon whiskers, from siloxene ($Si_6O_3H_6$) and so on.

To explain these effects, several mechanisms of PL were proposed: for example, the quantum confinement effect in microcrystalline structures; the surface effect of dihydride ($SiH_2$) passivation; the mechanism associated with $\alpha$-Si:H and some others. However, all these mechanisms are inconsistent with the observations among different research groups: the volume fraction of small clusters is not enough to produce the strong and narrow luminescence; at room temperature, the peak energy moves upwards with increasing excitation energy, but the spectral width of the luminescence spectra remains unchanged; a theoretical description of the strong luminescence in PS can be obtained by the extrapolation of conventional band structure calculations for Si: to the systems with dimensions d $\leq 3$ nm (and siloxene with d $\sim 0.5$ nm). However, the measurement with atomic force microscope shows the features mainly about 40 - 60 nm.

In this paper, the new model of the dipole emission of low dimension structures is proposed. The natural frequencies of them are determined by their dimensions and are located in the spectral range from infrared to ultraviolet. This model agrees well with our investigation of the optical parameters of the thin films of the series of semiconductors and metals and with the experiments on the luminescence in the different low dimension structures, for example, in PS.

**2. MICROSCOPIC APPROACH**

An index of refraction of light determines the effective velocity of light in a medium [7-9]. It is due to the fact that the incident light wave excites in a medium an emission of every atom. The total field in a medium $\vec{E}_T(\vec{r},t)$ is determined by the diffraction, interaction of the incident field $\vec{E}_i(\vec{r},t)$ and the field $\vec{E}_d(\vec{r},t)$ radiated by all dipoles of the medium:

$$\vec{E}_T(\vec{r},t) = \vec{E}_i(\vec{r},t) + \vec{E}_d(\vec{r},t)$$

We consider below the case of a normally incident linearly polarized monochromatic plane wave. The Ewald- Oseen extinction theorem states, that the part of the electromagnetic field radiated by the molecular dipoles, cancels the incident field and the other part propagates in accordance with the Maxwell equations and some parameters of a medium.

Ewald and several other researchers believe that the dipoles on the boundary compensate completely the incident wave inside the medium and create the dynamically self consistent mode of the propagated refracted wave and reflected wave as well [7-9] . At the same time, it is discussed question – other researchers (for example [10]), believe, that the effective dipole layer exists with the thickness $x \sim \dfrac{\lambda_0}{2\pi(n-1)}$, which produces the reflected wave.

We consider this problem from the other point of view: with the help of two independent approaches: micro- and macro- approaches, we find the dimension dependence of the optical parameters of very thin layers of materials ($d \ll \lambda_0$).

When a monochromatic plane wave falls on a surface of a crystal, the total field inside the layer with the thickness $d$, $\vec{E}_T(\vec{r},t)$, can be written in the form [11]:

$$E_T(z) = E_i^{(0)} \cdot \exp(ik_0 z) + \frac{i}{2} k_0 [\varepsilon(d) - 1] \cdot \{\exp(ik_0 z)\int_0^z dz' E(z') \cdot \exp(-ik_0 z') + \exp(-ik_0 z) \cdot \int_z^d dz' E(z') \cdot \exp(ik_0 z')\} \quad (1)$$

Here $\kappa_0 = \dfrac{\omega}{c} = \dfrac{2\pi}{\lambda_0}$ is a wave vector of light wave in a vacuum; $\varepsilon(d)$ is a dielectric function of a layer of a thickness $d$. The solution of the equation (1) will seek in the form: $E_T(z) = E_T^{(0)} \cdot \exp[ik(z) \cdot z]$. At a surface, $z = 0$, the ratio for the wave amplitudes, following from the Maxwell boundary conditions, takes place : $\dfrac{E_i^{(0)}}{E_T^{(0)}} = \dfrac{n(d)+1}{2}$. From the solution of the Maxwell equations in the case of "thick" layers ($d \gg \lambda_0$) it follows that the wave vector in a medium is equal to $k = k_0\sqrt{\varepsilon_\infty} = k_0 \cdot (n_\infty + i\kappa_\infty)$, where $\varepsilon_\infty$, $n_\infty, \kappa_\infty$ are a dielectric function, an index of refraction and index of extinction for "volume" material, respectively. Assume the same relation for a thin layers ($d \ll \lambda_0$) as well : $k(z) = k_0 \cdot [n(z) + i\kappa(z)]$. Introduce the dimensionless coordinates: $x = k_0 \cdot z; a = k_0 \cdot d$. In the case of a thin layer of a thickness $d$, $d \ll \lambda_0$, the integral equation can be written in the form:

$$\exp[-\kappa(a) \cdot a] \cdot \exp[i(n(a) - 1) \cdot a] = \frac{n(a) + 1}{2} +$$
$$+ \frac{i}{2} \cdot [[(n(a) + i\kappa(a))]^2 - 1] \cdot \int_0^a dx \cdot \exp[-\kappa(x) \cdot x] \cdot \exp[i[n(x) - 1] \cdot x] \quad (2)$$

To solve this integral equation, it is necessary to approximate the functions $n(x), \kappa(x)$. Let us use the model of smooth functions which in the limit $x \to 0$, trend to $n(x) \to 1$, $\kappa(x) \to 0$, respectively, and which smoothly saturate to $n_\infty, \kappa_\infty$ at $d \gg \lambda_0$:

$$n(x) - 1 \to (n_\infty - 1) \cdot [1 - \exp(-x)]$$
$$\kappa(x) \to \kappa_\infty \cdot [1 - \exp(-x)]$$

In the case of a nanometer thickness, the parameter $a = \frac{2\pi d}{\lambda_0} \ll 1$, hence, we can write approximately: $n(x) \approx 1 + [n_\infty - 1] \cdot x; \kappa(x) \approx \kappa_\infty \cdot x$. Dividing the real and imagine parts of the obtained equation, we put the system of two equations for $n(a)$ and $\kappa(a)$, i.e., the dimension dependences of two main optical parameters:

$$\exp[-\kappa(a) \cdot a] \cdot \cos[(n(a) - 1) \cdot a] = \frac{n(a) + 1}{2} - \frac{1}{2} \cdot [n^2(a) - \kappa^2(a) - 1] \cdot J_3 - n(a) \cdot \kappa(a) \cdot J_2; \quad (3)$$

$$\exp[-\kappa(a) \cdot a] \cdot \sin[(n(a) - 1) \cdot a] = \frac{1}{2} \cdot [n^2(a) - \kappa^2(a) - 1] \cdot J_2 - n(a) \cdot \kappa(a) \cdot J_3; \quad (4)$$

where 
$$J_2 = \int_0^a dx \cdot \exp[-\kappa_\infty x^2] \cdot \cos[(n_\infty - 1) \cdot x^2] \quad (5)$$

$$J_3 = \int_0^a dx \cdot \exp[-\kappa_\infty x^2] \cdot \sin[(n_\infty - 1) \cdot x^2] \quad (6)$$

If an extinction is absent, the integrals (5), (6) coincide with the Fresnel integers $S(a), C(a)$, respectively. It agrees with the fact, that diffraction of light on atoms in a medium, takes place.

Such a model microscopic calculation in the case of nanometer films, $a < 1$, leads to dimension dependences, whose qualitative forms present in fig.1. The maximum of the curve $n(a)$ and minimum of $\kappa(a)$ are seen, the value $\kappa(a)$ being negative, i.e. the emission takes place and not only absorption.

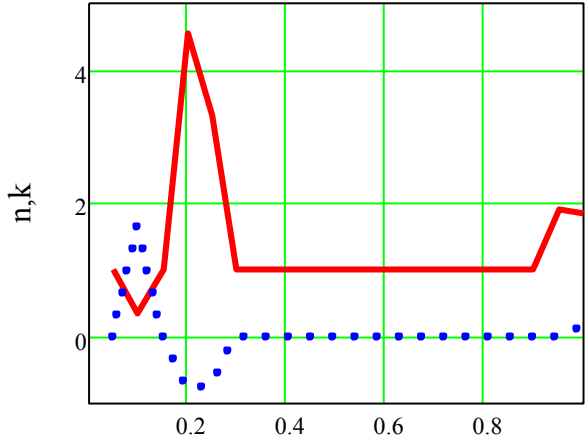

Fig.1. The qualitative dimension dependences of $n$ (solid) and $\kappa$ (dotted) according to the microscopic model approach in the limit of nanometer –scale thicknesses ($n_\infty = 3; \kappa_\infty = 0.2$).

### 3. MACROSCOPIC APPROACH

Now we carry out the calculation of $n(a)$ and $\kappa(a)$ on the base of geometrical optics, and the experimental data of different researchers for reflection and transmission of light in the case of nanometer-scale layers of semiconductors and metals [12-15].

To find an index of reflection $n$ and absorption $\kappa$ of isotropic materials, usually it is used one of the methods, based on experimental data of reflection, $\Re$, and transmission, $\Im$, (in the case of thin films) or only $\Re$ in the case of "volume" materials (for instance, [16]):

$$\Re = \frac{r_{12}^2 + r_{23}^2 \cdot \exp(-2\alpha d) - 2 r_{12} r_{23} \exp(-\alpha d) \cdot \cos\delta}{1 + r_{12}^2 r_{23}^2 \cdot \exp(-2\alpha d) - 2 r_{12} r_{23} \exp(-\alpha d) \cdot \cos\delta} \quad (7)$$

$$\Im = \frac{T_{12} T_{23} \cdot \exp(-\alpha d)}{1 + r_{12}^2 r_{23}^2 \cdot \exp(-2\alpha d) - 2 r_{12} r_{23} \exp(-\alpha d) \cdot \cos\delta} \quad (8)$$

The formulas for $\Re'$ differs from $\Re$ by the index change $1 \Leftrightarrow 3$.

For the calculation of $n$ and $\kappa$, the Fresnel formulas for the wave amplitudes and the Snell law of refraction are usually used, which were derived for the transparent media with the substitution $n \to n + i \cdot \kappa$ for the account of absorption. The proof of this change was never made. It was customary to think that, since the Maxwell boundary conditions are valid both for transparent and absorbing media, and, on the other hand, the solution of the wave equation for the electric field in absorbing media differs from the solution for the electric field in transparent media by the substitution $n \to n + i \cdot \kappa$, then any relation of optics for transparent media can be formally generalized to the case of absorbing media by the same substitution. But it is not the case. Such a formal unproved approach leads to the series difficulties. For example, in the Snell law, the angle of refraction becomes complex and loses the physical meaning of angle: $\sin\vartheta_2 = \sin\vartheta_1 \cdot \frac{n_1}{(n_2 + i k_2)}$; the calculation of $n, \kappa$ led to nonphysical results [2,14, 17-19]: when the thickness of a film is decreased until it becomes almost transparent, the index of refraction increased sharply. At the same time, the measurement of $n(d)$ for thin films in the exciton absorption region by the prism method, has shown that $n(d)$ decreases when $d$ decreases and $n(d) \to 1$, when $d \to 0$.

Therefore we have derived the formulas for the wave amplitudes for absorbing and conducting media from the Maxwell boundary conditions. The absorption and conductance were directly taking into account in the boundary conditions [20,21,1].

The obtained formulas for the wave amplitudes and the laws of refraction differ substantially from the formulas, written for absorbing and conducting media with the formal substitution $n \to n + i \cdot \kappa$. It is shown that the laws of refraction are different for the TE- and the TM- wave. In the case of the TE-wave, the form of the law of refraction coincides with Snell's law for transparent media and only traditional (positive) refraction takes place. In the case of the TM-wave, the law of refraction has a more complicated form. In the limit case of transparent media, it coincides with Snell's law; in the limit case of the exciton absorption, it coincides with our former results for excitons; and in the limit case of a medium with a very high conductivity, the value of the sine of the angle of refraction becomes negative, $\sin\theta_2 < 0$, i.e. negative refraction takes place (with all positive optical parameters). This phenomenon is due to the surface current. In good conductors, this effect can be observed up to the ultraviolet region of the spectrum: in Ag, Cu and Au. In Al and other metals and in all semiconductors, this effect can only be observed in the infrared region. Our conclusion agrees well with the experiments on different metamaterials. Negative refraction was observed in a thin silver slab in the UV region of the spectrum and in three-dimensional semiconductor structures in the infrared region of the spectrum.[23,24]

Let the plane linearly polarized monochromatic plane wave of the frequency $\omega$, fall on the boundary of two media: the first – with an index of refraction $n_1$ and absorption $\kappa_1$, and the second - with an index of refraction $n_2$ and absorption $\kappa_2$. In the boundary conditions, one must take into account the fact, that in absorbing medium, the electric induction has two components: the component in phase with the electric field of the incident wave and another one – out of phase by $\frac{\pi}{2}$. Due to this fact the dielectric function of an absorbing medium is complex. The form of these equations in general case and some special cases are given in Appendix A, B. In Appendix C, it is shown, that in the case of absorbing and conducting media, the law of energy conservation is valid, both for TM- and TE-polarizations of light.

## 4. DETERMINATION OF THE OPTICAL PARAMETERS

Now let us carry out the calculation of the dimension dependencies of an index of refraction and absorption, using the formulas (7),(8) and the formulas, obtained above for the wave amplitudes.

Suppose the normal incidence of monochromatic wave light on a plane parallel plate, with the wavelength in a vacuum $\lambda$. The coefficient of an absorption of material is equal to $\alpha = \frac{4\pi\kappa}{\lambda}$. $r_{ij}, t_{ij}$ - are the ratios of the amplitudes – reflected to incident and refracted to incident, respectively. $T_{ij} = t_{ij} \cdot t_{ji}$ is a transparence. In the case of a normal incidence, the difference in phase between every $s-$ and $s+1-$ term of a series of refracted and reflected waves at a manifold interference is equal to: $\delta = \frac{4\pi}{\lambda} n_2 d$. Above it was shown, that at normal incidence the formulas for $r_{ij}, t_{ij}$ coincide in a form with those, which were in the case of transparent media, but the sense of them is another (they are determined by conductance and not polarizability):

$$r_{ij} = \frac{n_j - n_i}{n_j + n_i}; t_{ij} = \frac{2n_i}{n_j + n_i}$$

The meaning of $n_2, \kappa_2$ can be calculated, when one of the pairs $\mathfrak{R}, \mathfrak{I}$, or $\mathfrak{R}', \mathfrak{I}$, or $\mathfrak{R}, \mathfrak{R}'$ are measured. For the control of correctness of our calculations, the optical parameters of the investigated materials were calculated both with experimental data of the pair $\mathfrak{R}, \mathfrak{I}$, and of $\mathfrak{R}, \mathfrak{R}'$.

The numerical calculation of the dimension dependences of $n(d)$ and $\kappa(d)$ for certain materials were carried out at fixed wavelength $\lambda$, a crystal thickness $d$, and indices of refraction of transparent media $n_1, n_3$, according to the experimental conditions of the used experimental works, where $\mathfrak{R}, \mathfrak{I}$; $\mathfrak{R}, \mathfrak{R}'$ have been measured. The calculations were carried out with the experimental data of [2], where reflectance and transmittance of thin polycrystalline layers of $Ge, Se, Si, Te$ were measured; with the data of [12,13], where reflectance and transmittance of thin films of silver were measured; with the data of [14], where reflectance and transmittance of thin films of aluminum were measured; with the data of [15], where reflectance and transmittance of thin films of iron were measured. In all these works, the measurements of a film thickness were made with independent methods. Some of our calculations of $n(d)$ have been present in [1].

The results of our calculations of $n(d)$ and $\kappa(d)$ are present on fig.2-6.

In all investigated materials at small thicknesses, the function $n(d)$ decreases and when $d \to 0$, then $n(d) \to 1$ (as one can expects). At the thicknesses of a film $d \to 80-100$ нм, the meanings of $n(d)$ trend to their "volume" values. It is seen, that calculations of $n(d)$ with the experimental data of the pairs $\mathfrak{R}, \mathfrak{I}$ and $\mathfrak{R}, \mathfrak{R}'$ (circles and triangles (in figures 2,4,6) , respectively), lead to almost the same meanings of $n(d)$ and $\kappa(d)$. It shows the correctness of our formulas for the wave amplitudes.

In a region of nanometer-scale thicknesses (15 -50 nm for different materials and different wavelengths) all obtained curves $n(d)$ and $\kappa(d)$ have the thickness resonance - maxima of $n(d)$ and minima of $\kappa(d)$, which corresponds to the diffraction conditions at the wavelength of light on a medium, $\lambda_m = \frac{\lambda}{n(d)} : d_{res} = \frac{\lambda}{\pi n(d_{res})} \cdot m; (m = 1,2,3..)$.

On the circle with a radius $\frac{d}{2}$, the integer number of wavelength, $m$, are placed. It takes place for a resonator, when light emission occurs. In semiconductors, only $m = 1$ takes place, in the same time, in metals, (for example, in Ag, Al, Au) the resonance occurs for $m = 1,2,3$ (fig.4). The curves $\kappa(d)$ have minimum, hence, both absorption and emission takes place.

The films of iron have much less conductivity in comparison with silver film, and only one resonance is observed, $m = 1$ (fig.6).

In all investigated materials, the index of refraction is more than unity, $n > 1$. This agrees with the fact that polarization of a medium is always positive.

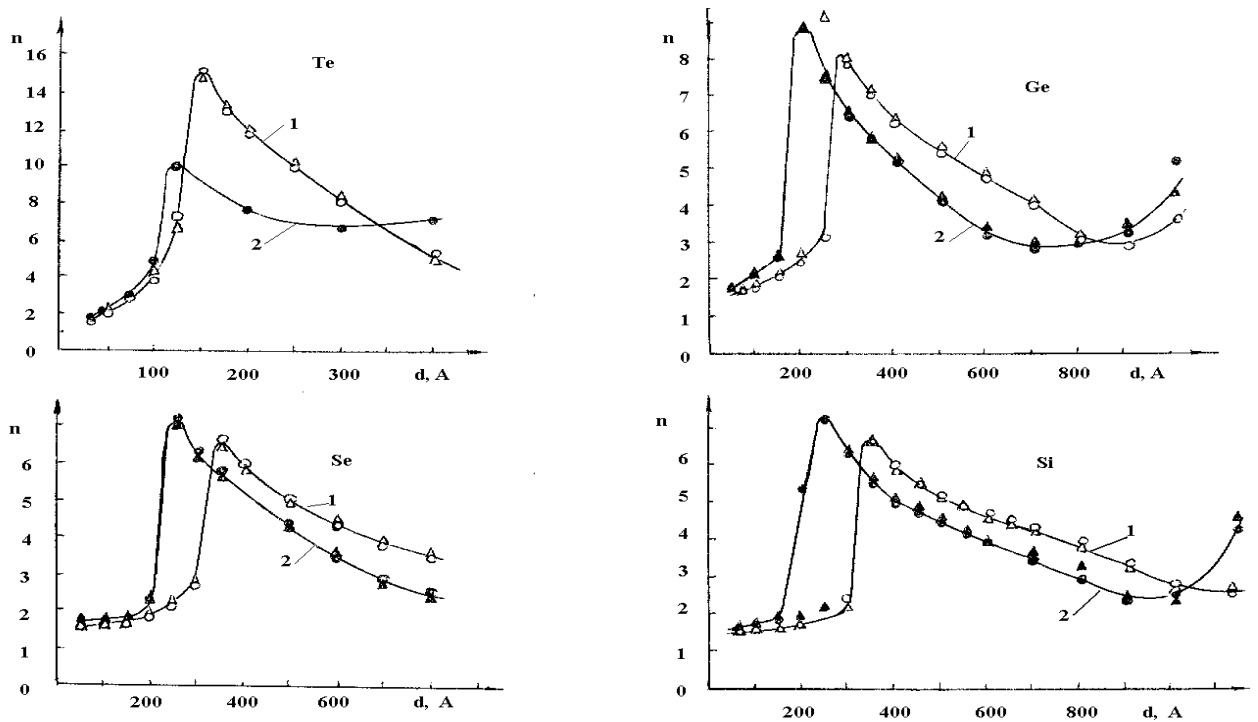

Fig.2. Calculated dimension dependences of $n(d)$ in semiconductors $Ge, Se, Si, Te$ (for data of [2]). The curves 1 and 2 correspond to the wavelengths $\lambda = 700$ nm and 560 nm, respectively.

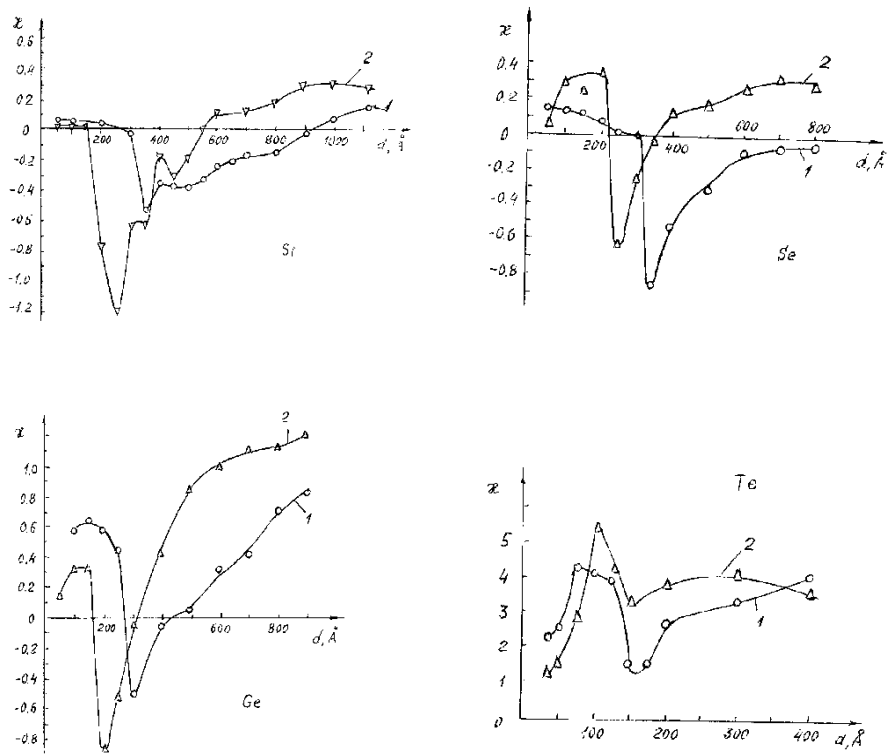

Fig.3. The calculated curves $\kappa(d)$ (the same materials and conditions as in fig.2).

The coefficient of absorption $\alpha$ is equal to: $\alpha = \dfrac{4\pi\kappa}{\lambda}$. The calculations show, that in thin films of even very good metals (Ag, Al) the effective index of refraction does not equal to the effective index of absorption: $n^* \neq \kappa^*$.

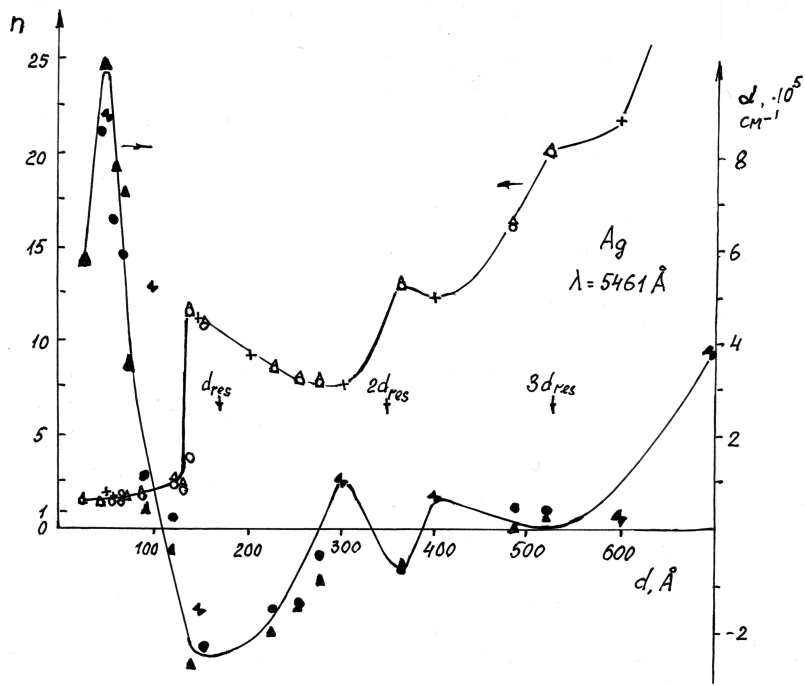

Fig.4. Calculated dimension dependences of $n(d)$ and the coefficient of absorption $\alpha(d)$ in thin layers of silver (for data of [12], circles, and of [13], crosses). ($\lambda$ = 546.1 nm).

In all investigated metals and semiconductors in some region of nanometer-scale thicknesses, not only absorption takes place, but emission of light as well. The measured coefficient of absorption, $\alpha = \alpha_{abs} - \beta_{emis}$, becomes negative due to predominance of the coefficient of emission, $\beta_{emis}$.

A conductivity in metal can be calculated according to its index of refraction and absorption:

$$\sigma = \frac{n\kappa \cdot \omega}{2\pi} = \frac{\alpha c n}{4\pi}$$

In thick crystals of silver, $d > 100$ nm, and wavelength of light $\lambda = 546.1$ nm, the coefficient of absorption is near $\alpha \sim 10^6 cm^{-1}$, and $n \approx 35$. Estimate the conductivity at this frequency ($\omega = 3.45 \cdot 10^{15} cek^{-1}$): $\sigma \approx 8.3 \cdot 10^{16} cek^{-1}$. At low frequencies the silver conductivity is equal: $\sigma \approx 5.42 \cdot 10^{17} cek^{-1}$, it is less in 6.53 times in the visible region of spectrum. In the table 1, the values of conductivity of *Ag, Al* and *Fe* for these two frequencies are given.

## 5. COMPERISON WITH THE MICRO-CALCULATION

Macro- and micro- approaches to the determination of the optical parameters of nanometer-scale layers show that their properties differ considerably from the "volume" ones ($d >> \lambda_0$). Both considerations show that: an index of refraction and absorption of nanometer-scale layers depend strongly on a film thickness; there are the regions of dimension resonance, where functions $n(d)$ have the maximum, and $\kappa(d)$ have minimum, and the values of $\kappa(d)$ reaching the negative meanings.

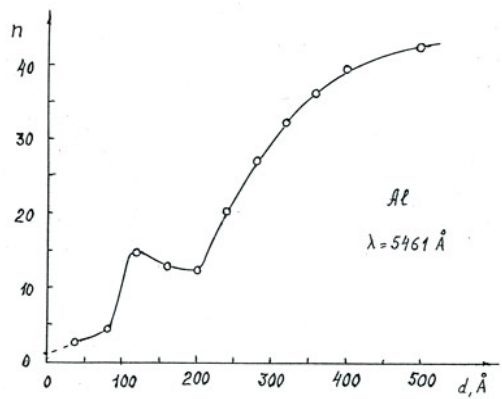

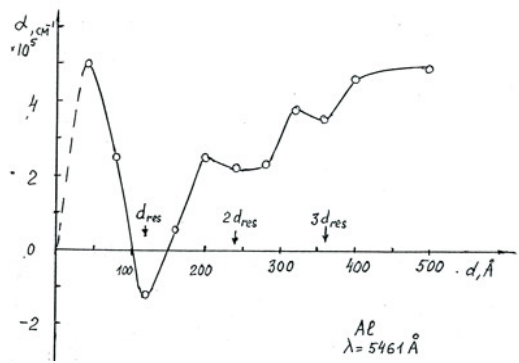

Fig.5. Calculated dimension dependences of $n(d)$ and the coefficient of absorption $\alpha(d)$ in thin layers of aluminum (for data of [14]). ($\lambda = 546.1$ nm).

Micro-calculation were not carried out for a certain material, only some value of the "volume" ($d \gg \lambda_0$) optical parameters were supposed. Therefore this approach is only qualitative. Still it is important, that both considerations give similar results.

Regarding the question - is there an extinction theorem distance, and the role of surface dipoles, macro-calculation indicates that besides the boundary itself, in nanometer-scale layers, the considerable contribution of the many-fold interference into the reflected and refracted waves, takes place, i.e. the second boundary, absorption of light in a layer and its reemission, which is determined by the resonator (the layer itself). The layer with the thickness $x \sim \dfrac{\lambda_0}{2\pi(n-1)}$ is a limit for "volume" material, when the effects of many-fold interference are absent.

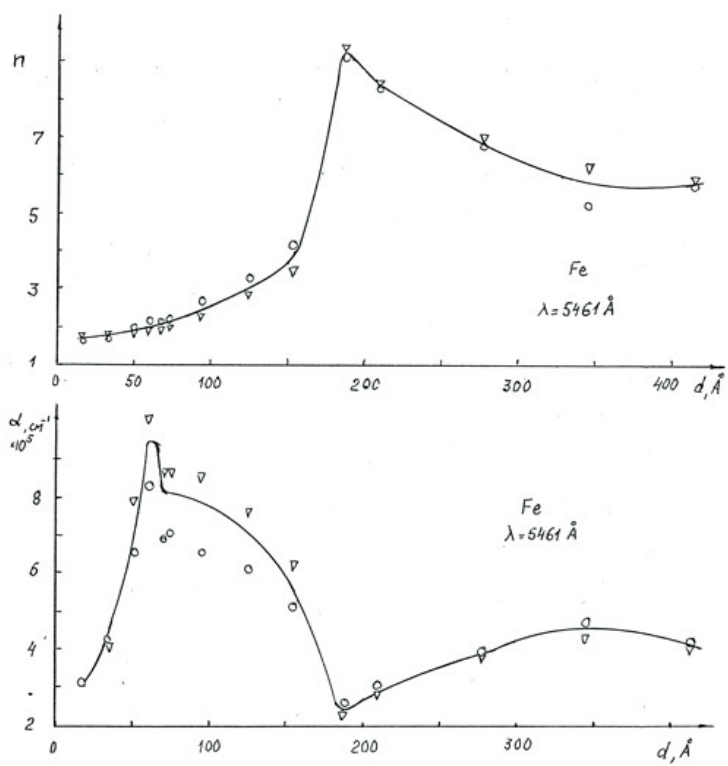

Fig.6. . Calculated dimension dependences of $n(d)$ and the coefficient of absorption $\alpha(d)$ in thin layers of iron (for data of [15, circles- on $\Re, \Im$ and triangles- on $\Re, \Re'$). ($\lambda = 546.1$ nm).

Table 1

|  | Ag | Al | Fe |
|---|---|---|---|
| $\sigma_{lowfr} \cdot 10^{17} cek^{-1}$, low freq. | 5.42 | 3.17 | 0.875 |
| $\sigma_{light} \cdot 10^{16} cek^{-1}$, $\lambda = 546.1$ nm | 8.3 | 4.8 | 0.57 |
| $\dfrac{\sigma_{lowfr}}{\sigma_{light}}$ | 6.48 | 6.6 | 15.2 |

## 6. CONCLUSION

The investigation of the optical properties of thin nanometer-scale layers of materials shows that both an index of refraction and coefficient of absorption depend strongly on a their thickness. In this region of thicknesses, the dimension resonance occurs, where an index of refraction has a maximum and a coefficient of absorption has a minimum. The reemission is possible for the thicknesses, which is multiple to the wavelength of light in a medium: $\dfrac{2\pi d_{res}}{2} = \lambda_m \cdot m = \dfrac{\lambda_0}{n} \cdot m$, where $m = 1,2,3\ldots$ The calculations show the existence of these extremes in all investigated materials: *Ag, Al, Fe, Ge, Si, Se, Te*.

The calculation of the optical parameters have been carried out using the experimental data of reflection and transparency of thin layers, obtained in a series of works, and our formulas for the wave amplitudes and the laws of refractions. The analogues of the Fresnel formulas and the Snell law have been derived from the Maxwell boundary conditions where the absorption and conductivity of media were taken into account. The derived formulas differ substantially from those which were obtaining by the formal substitution $n \rightarrow n + i \cdot \kappa$ in all formulas derived for transparent media.

It is shown that invoking our formulas leads to fulfillment of the conservation law of energy both for the TE- and TM- polarizations of light. The obtained dimension dependences of the optical parameters of nanometer-scale layers of semiconductors and metals can be useful for practical applications.

The visible luminescence of the porous silicon, silicon whiskers and other nanometer – scale silicon structures can be explained by their emission in the case of a dimension resonance because nanometer –scale silicon layer has a great negative coefficient of absorption and a little (in comparison with Ag) damping.

**Appendix A. BOUNDARY CONDITIONS. THE GENERAL CASE.**

The Maxwell boundary conditions in general case have the form:

$$\vec{n}_{12} \cdot \{\sum_k \sqrt{\mu_{2k}} \cdot n_{2k}[\vec{s}_{2k} \times \vec{E}_{2k}] - \sum_j \sqrt{\mu_{1j}} \cdot n_{1j}[\vec{s}_{1j} \times \vec{E}_{1j}]\} = 0$$

$$\vec{n}_{12} \cdot \{\sum_k \sqrt{\mu_{2k}} \cdot \kappa_{2k}[\vec{s}_{2k} \times \vec{E}_{2k}] - \sum_j \sqrt{\mu_{1j}} \cdot \kappa_{1j}[\vec{s}_{1j} \times \vec{E}_{1j}]\} = 0$$

$$[\vec{n}_{12} \times \{\sum_k \vec{E}_{2k} - \sum_j \vec{E}_{1j}\}] = 0$$

$$\vec{n}_{12} \cdot \{\sum_k (n_{2k}^2 - \kappa_{2k}^2) \cdot \vec{E}_{2k} - \sum_j (n_{1j}^2 - \kappa_{1j}^2) \cdot \vec{E}_{1j}\} = 4\pi \rho_1 \qquad (A1)$$

$$\vec{n}_{12} \cdot \{\sum_k n_{2k}\kappa_{2k} \cdot \vec{E}_{2k} - \sum_j n_{1j}\kappa_{1j} \cdot \vec{E}_{1j}\} = 2\pi \rho_2$$

$$\{\sum_k n_{2k} \cdot [\vec{n}_{12} \times [\vec{s}_{2k} \times \vec{E}_{2k}]] - \sum_j n_{1j} \cdot [\vec{n}_{12} \times [\vec{s}_{1j} \times \vec{E}_{1j}]]\} = \frac{4\pi}{c} \cdot j_1$$

$$\{\sum_k \kappa_{2k} \cdot [\vec{n}_{12} \times [\vec{s}_{2k} \times \vec{E}_{2k}]] - \sum_j \kappa_{1j} \cdot [\vec{n}_{12} \times [\vec{s}_{1j} \times \vec{E}_{1j}]]\} = \frac{4\pi}{c} \cdot j_2$$

Here $\vec{E}$ and $\vec{H}$ - are the electric and magnetic field vectors of the electromagnetic wave; $\vec{D}$ and $\vec{B}$ - are the electric and magnetic induction vectors; $\rho$ and $\vec{j}$ – the surface charge and current density; $\varepsilon$ and $\mu$ - are the dielectric function and the magnetic susceptibility of a medium; $\sigma$ is the conductance of a medium. Here we use the second Maxwell equation and obtain the relation between $\vec{E}$ and $\vec{H}$: $\vec{H} = (n + i\kappa) \cdot [\vec{s} \times \vec{E}]$. $\vec{s}$ – is the unit vector in the direction of the wave propagation. Below we put $\mu = 1$.

Now several refracted waves can exist. The oscillating charges and currents can be present in a form of two component as well:

$$\hat{\rho} = (\rho_1 + i \cdot \rho_2) \cdot \exp(i\omega t)$$

$$\vec{\hat{j}} = (\vec{j}_1 + i \cdot \vec{j}_2) \cdot \exp(i\omega t) \qquad (A2)$$

In the case of a transparent media, the system (A1) reduces to the system of three equation:

in the case of the TM-polarization:

$$E_{2t} \cdot \cos\theta_2 + (E_{1R} - E_i) \cdot \cos\theta_1 = 0$$

$$E_{2t} \cdot n_2^2 \sin\theta_2 - n_1^2 (E_{1R} + E_i) \cdot \sin\theta_1 = 0 \qquad (A3)$$

$$E_{2t} \cdot n_2 - n_1 (E_{1R} + E_i) = 0$$

Here $E_R, E_t, E_i$ - are the amplitudes of reflected, refracted and incident waves.

The system (A3) gives the Fresnel formulas for the wave amplitudes:

$$E_{2t}^{(p)} = E_i^{(p)} \cdot \frac{2n_1 \cos\theta_1}{n_1 \cos\theta_2 + n_2 \cos\theta_1}$$

$$E_{1R}^{(p)} = E_i^{(p)} \cdot \frac{n_2 \cos\theta_1 - n_1 \cos\theta_2}{n_1 \cos\theta_2 + n_2 \cos\theta_1}$$

And the Snell law of refraction: $\sin\theta_2 = \frac{n_1}{n_2} \sin\theta_1$.

In the case of the TE-polarization:

$$E_{2t} \cdot n_2 \cos\theta_2 + (E_{1R} - E_i) n_1 \cos\theta_1 = 0$$

$$E_{2t} - (E_{1R} + E_i) = 0$$

$$E_{2t} \cdot n_2 \sin\theta_2 - n_1 (E_{1R} + E_i) \cdot \sin\theta_1 = 0$$

and the Fresnel formulas:

$$E_{2t}^{(s)} = E_i^{(s)} \cdot \frac{2n_1 \cos\theta_1}{n_1 \cos\theta_1 + n_2 \cos\theta_2}$$

$$E_{1R}^{(s)} = E_i^{(s)} \cdot \frac{n_1 \cos\theta_1 - n_2 \cos\theta_2}{n_1 \cos\theta_1 + n_2 \cos\theta_2}$$

In the case of a transparent medium, the law of refraction does not depend upon polarization of light.

In the case of the exciton, plasmas, lattice absorption, a charge and current are equal to zero, but absorption takes place. This case has been formerly investigated in detail in our works [20- 22]. From the Maxwell boundary conditions we have the systems of five equations both for TE- and TM-polarizations and the existence of three refracted waves is simultaneously possible (which were found experimentally as well [21,22] ). The condition of consistent equations is the equality to zero of the corresponding fifth-order determinant; it gives the law of refraction for every refracted to the second medium wave. The law of refraction of the TM-wave:

$$\sin\theta_{2k}^{(p)} = \frac{n_1 n_{2k}}{n_{2k}^2 - \kappa_{2k}^2} \cdot \sin\theta_1; \qquad k = 1,2,3.$$

For the TE-wave: $\sin\theta_{2k}^{(s)} = \frac{n_1}{n_{2k}} \sin\theta_1 \quad ; k = 1,2,3.$

In the case of TE-polarization the law of refraction coincides with the Snell law in the form, but now –for each of three wave. The wave amplitudes are obtained after the solution of other four equations of the systems, They are the analogous of the Fresnel formulas for the case of the exciton absorption.

**Appendix B. BOUNDARY CONDITIONS. THE ABSORBING AND CONDUCTING MEDIA.**

If the medium is absorbing, but its conductivity is very few, then the formulas for the wave amplitudes and the laws of refraction are the same as in the case of exciton absorption, but for one wave only.

Let us consider the case when the first medium is transparent, $n_1$ and $\hat{\rho}_{(1)} = \hat{j}_{(1)} = 0, \sigma_{(1)} = 0$, the second – is absorbing and conducting: $n_2, \kappa_2$ and $\sigma_{(2)}$; $\hat{\rho}_{(2)} \neq 0; \hat{j}_{(2)} \neq 0$. In this case from the Maxwell boundary conditions we have systems of five equations for both polarization as well.

Electromagnetic wave in the second medium has the form:

$$\vec{E}(\vec{r},t) = \vec{E}_0 \cdot \exp[-i(\omega t - \tilde{k}(\vec{r}\cdot\vec{s}))]$$

here $\vec{s}$ is the unit vector in the direction of the wave propagation; $\tilde{k} = \frac{\omega}{c}(n^* + i\kappa^*)$

$$n^* = \frac{1}{\sqrt{2}} \cdot \{\sqrt{\mu^2\varepsilon^2 + (\frac{4\mu\pi\sigma}{\omega})^2} + \mu\varepsilon\}^{1/2}$$

$$\kappa^* = \frac{1}{\sqrt{2}} \cdot \{\sqrt{\mu^2\varepsilon^2 + (\frac{4\mu\pi\sigma}{\omega})^2} - \mu\varepsilon\}^{1/2}$$

In good conductors both an index of refraction and index of absorption are mainly determined by a conductivity. At frequencies below the plasmas resonance frequency ($\omega < \omega_p$, $\omega_p$ – is a frequency of plasmas resonance), when $\mu\varepsilon << \frac{4\mu\pi\sigma}{\omega}$, then approximately we can write:

$$n^* \approx \sqrt{\frac{2\pi\sigma}{\omega}} \cdot [1 + \frac{\varepsilon\omega}{8\pi\sigma}],$$

$$\kappa^* \approx \sqrt{\frac{2\pi\sigma}{\omega}} \cdot [1 - \frac{\varepsilon\omega}{8\pi\sigma}]$$

For the TM-polarization in this case we have a system of five equations:

$$E_t \cdot \cos\theta_2 + (E_R - E_i) \cdot \cos\theta_1 = 0$$

$$E_t \cdot n_2^2 \cdot \sin\theta_2 - n_1^2(E_R + E_i) \cdot \sin\theta_1 = 4\pi\hat{\rho}_1$$

$$E_t \cdot n_2\kappa_2 \sin\theta_2 = 2\pi\hat{\rho}_2 \approx 0$$

$$E_t \cdot n_2^* - n_1(E_R + E_i) = -\frac{4\pi}{c}\hat{j}_1$$

$$E_t \cdot \kappa_2^* = -\frac{4\pi}{c}\hat{j}_2$$

Here the amplitudes of the oscillating surface charge and current is equal:

$$\hat{\rho} = \hat{\rho}_1 + i\hat{\rho}_2$$

$$\hat{j} = \hat{j}_1 + i\hat{j}_2$$

The current component $\hat{j}_2$ determined by the last equation of the system. At first, for the component $\hat{j}_1$ we introduce the relation: $\hat{j}_1 = \hat{\sigma}_2 \cdot E_t$, where $\hat{\sigma}_2$ is a surface conductivity. Below we obtain its meaning as a function of $n^*, \kappa^*$.

The surface charge component $\hat{\rho}_2$ is determined by the third equation of the system, and at first, for the component $\hat{\rho}_1$ we introduce the relation $4\pi\hat{\rho}_1 = \beta \cdot \sin\theta_2 \cdot E_t$. Below we find $\beta$ as a function of $n^*, \kappa^*$.

The law of refraction for TM-wave we obtain from the second and forth equations of the system:

$$\sin\theta_2^{(p)} = \frac{n_1(n_2^* + \frac{4\pi}{c}\hat{\sigma}_2)}{n_2^2 - \beta} \cdot \sin\theta_1^{(p)}$$

For the amplitudes of refracted and reflected waves we obtain:

$$E_t^{(p)} = E_i^{(p)} \cdot \frac{2n_1\cos\theta_1}{n_1\cos\theta_2 + (n_2^* + \frac{4\pi}{c}\hat{\sigma}_2)\cos\theta_1}$$

$$E_R^{(p)} = E_i^{(p)} \cdot \frac{(n_2^* + \frac{4\pi}{c}\hat{\sigma}_2)\cos\theta_1 - n_1\cos\theta_2}{n_1\cos\theta_2 + (n_2^* + \frac{4\pi}{c}\hat{\sigma}_2)\cos\theta_1}$$

The parameters $\beta, \hat{\sigma}_2$ can be found with the use of the equation of continuity;

$$\frac{\partial\hat{\rho}}{\partial t} = -div\hat{\vec{j}}$$

In the case of harmonic wave: $-i\omega\hat{\rho} = -i\tilde{k}_x\hat{j}$, hence, we have:

$$\hat{\rho}_1 = \frac{1}{c}(n_2^*\hat{j}_1 - \kappa_2^*\hat{j}_2)\sin\theta_2 = \frac{\beta}{4\pi}\sin\theta_2 E_t$$

$$\hat{\rho}_2 = \frac{1}{c}(n_2^*\hat{j}_2 + \kappa_2^*\hat{j}_1)\sin\theta_2 \approx 0$$

The solution gives:

$$\hat{\sigma}_2 = \frac{c}{4\pi} \cdot n_2^*$$

$$\beta = (n_2^*)^2 + (\kappa_2^*)^2$$

Now, the law of refraction for the TM-wave has the form:

$$\sin\theta_2^{(p)} = \frac{n_1 \cdot 2n_2^*}{n_2^2 - [(n_2^*)^2 + (\kappa_2^*)^2]} \cdot \sin\theta_1^{(p)}$$

It is seen, that in the case of a good metal and frequencies $\omega < \omega_p$,

$$\sin\theta_2^{(p)} \approx -\frac{n_1}{n_2^*} \cdot \sin\theta_1^{(p)} < 0$$

In this case, the angle of refraction is determined by the surface current:

$$\sin\theta_2^{(p)} \approx -n_1 \cdot \sqrt{\frac{\omega}{2\pi\sigma}} \cdot \sin\theta_1^{(p)}.$$

It means, that light refracts in a "wrong" direction (at all positive optical parameters).

In the case of the TM- polarization a surface current has a form:

$$\widehat{j}_{(p)} = \frac{c}{4\pi}[(n_2^* - i\kappa_2^*)] \cdot E_t$$

In the case of very high conductivity:

$$\widehat{j}_{(p)} \approx \frac{c}{4\pi} \cdot n_2^* \cdot E_t \cdot \exp(-\frac{i\pi}{4}) = c\sqrt{\frac{\sigma}{8\pi\omega}} \cdot E_t \cdot \exp(-\frac{i\pi}{4})$$

For the TE-polarization in this case we have a system of five equations:

$$E_t \cdot n_2^* \sin\theta_2 - n_1(E_R + E_i) \cdot \sin\theta_1 = 0$$

$$E_t \cdot \kappa_2^* \sin\theta_2 = 0$$

$$E_t - (E_R + E_i) = 0$$

$$E_t \cdot n_2^* \cos\theta_2 + (E_R - E_i)n_1\cos\theta_1 = -\frac{4\pi}{c}\widehat{j}_1$$

$$E_t \cdot \kappa_2^* \cos\theta_2 = -\frac{4\pi}{c}\widehat{j}_2$$

The law of refraction for TE-wave we obtain from the first and third equations of the system, it is similar to the Snell law, but the value $n_2^*$ has another sense:

$$\sin\theta_2^{(s)} = \frac{n_1}{n_2^*}\sin\theta_1$$

The refracted wave of TE-polarization has always "right" direction:

$$\sin\theta_2^{(s)} \geq 0$$

In the case of TE-polarization a surface current has always only one component: $\hat{j}_2^{(s)} = 0$. The total surface current is equal to:

$$\hat{j}^{(s)} = \hat{j}_1^{(s)} = \hat{\sigma}_2 E_t$$

$$\hat{j}^{(s)} \approx \frac{c}{4\pi} \cdot n_2^* \cdot E_t = c\sqrt{\frac{\sigma}{8\pi\omega}} \cdot E_t.$$

The amplitudes of the refracted and reflected waves is equal:

$$E_t^{(s)} = E_i^{(s)} \cdot \frac{2n_1 \cos\theta_1}{n_1 \cos\theta_1 + n_2^* \cos\theta_2 + \dfrac{4\pi}{c}\hat{\sigma}_2}$$

$$E_R^{(s)} = E_i^{(s)} \cdot \frac{n_1 \cos\theta_1 - n_2^* \cos\theta_2 - \dfrac{4\pi}{c}\hat{\sigma}_2}{n_1 \cos\theta_1 + n_2^* \cos\theta_2 + \dfrac{4\pi}{c}\hat{\sigma}_2}$$

In the limit of normal incidence, the amplitudes of refracted waves, $E_t^{(s)}, E_t^{(p)}$, coincides, and the amplitudes of reflected waves, $E_R^{(s)}, E_R^{(p)}$, have the opposite sign, just the same, as in the case of transparent media.

**Appendix 3. REFLECTIVITY AND TRANSPARENCY OF MEDIA**

Let show, that in the case of absorbing media, the law of energy conservation is valid, both for TM- and TE-polarizations of light.

Poynting's vector determines the amount of energy, flowing in unit of time, through the base of cylinder whose axis is parallel to $\vec{s}$ and the square of cross section is equal to unity, and is equal to: $\vec{S} = \frac{c}{4\pi}[\vec{E} \times \vec{H}] = \frac{c}{4\pi} EH \cdot \vec{s}$. The light intensity is equal: $S = \frac{c}{4\pi}\sqrt{\varepsilon}E^2 = \frac{cn}{4\pi}E^2$. The amount of energy in incident wave per unit of square in 1 sec, is equal to:

$$J^{(i)} = S^{(i)} \cdot \cos\vartheta_1 = \frac{cn_1}{4\pi} \cdot |E_i|^2 \cos\vartheta_1$$

The amount of energy in reflected and refracted waves per unit of square in 1 sec, is equal to:

$$J^{(r)} = S^{(r)} \cdot \cos\vartheta_1 = \frac{cn_1}{4\pi} \cdot |E_R|^2 \cos\vartheta_1$$

$$J^{(t)} = S^{(t)} \cdot \cos\vartheta_2 = \frac{cn_2}{4\pi} \cdot |E_t|^2 \cos\vartheta_2$$

Reflectivity and transparency are equal, respectively, to:

$$\mathfrak{R} = \frac{J^{(r)}}{J^{(i)}} = \left|\frac{E_R}{E_i}\right|^2$$

$$\mathfrak{I} = \frac{J^{(t)}}{J^{(i)}} = \frac{n_2 \cos\vartheta_2}{n_1 \cos\vartheta_1}\left|\frac{E_t}{E_i}\right|^2$$

For transparent media: $\mathfrak{R} + \mathfrak{I} = 1$.

In the general case, when an absorption, currents and charges in a medium take place, the law of energy conservation in the stationary case, has the form:

$$\int_V \vec{j}\cdot\vec{E}\,dV + \frac{c}{4\pi}\int_S [\vec{E}\times\vec{H}]\cdot\vec{n}\,ds = 0$$

The amounts of energy in an incident, reflected and refracted waves per unit of square in 1 sec, are determined by the formulas for $J^{(i)}, J^{(r)}, J^{(t)}$. The amount of energy in a surface current is equal to:

$$I^{(j)} = \hat{j}_d \cdot E_t \cdot \cos\theta_2$$

In the case of TE-polarization these expressions are equal:

$$J^{(i)}_{(s)} = \frac{cn_1}{4\pi}|E_{is}|^2\cos\theta_1$$

$$J^{(R)}_{(s)} = \frac{cn_1}{4\pi}|E_{is}|^2\cos\theta_1 \cdot \frac{(n_1\cos\theta_1 - n_2^*\cos\theta_2 - \frac{4\pi}{c}\hat{\sigma}_2)^2}{(n_1\cos\theta_1 + n_2^*\cos\theta_2 + \frac{4\pi}{c}\hat{\sigma}_2)^2}$$

$$J^{(t)}_{(s)} = \frac{cn_2^*}{4\pi}|E_{is}|^2\cos\theta_2 \cdot \frac{4n_1^2\cos^2\theta_1}{(n_1\cos\theta_1 + n_2^*\cos\theta_2 + \frac{4\pi}{c}\hat{\sigma}_2)^2}$$

$$I^{(j)}_{(s)} = \hat{\sigma}_2 \cdot |E_{is}|^2 \cdot \cos\theta_2 \cdot \frac{4n_1^2\cos^2\theta_1}{(n_1\cos\theta_1 + n_2^*\cos\theta_2 + \frac{4\pi}{c}\hat{\sigma}_2)^2}$$

The reflectivity is equal:

$$R_{(s)} = \frac{(n_1\cos\theta_1 - n_2^*\cos\theta_2 - \frac{4\pi}{c}\hat{\sigma}_2)^2}{(n_1\cos\theta_1 + n_2^*\cos\theta_2 + \frac{4\pi}{c}\hat{\sigma}_2)^2}$$

The transparency is equal: $\quad T_{(s)} = \dfrac{4n_1 n_2^* \cos\theta_1 \cos\theta_2}{(n_1\cos\theta_1 + n_2^*\cos\theta_2 + \frac{4\pi}{c}\hat{\sigma}_2)^2}$

The energy of the surface current is equal:

$$I_{(s)}^{(j)} = \frac{\frac{16\pi}{c} n_1 \hat{\sigma}_2 \cos\theta_1 \cos\theta_2}{(n_1 \cos\theta_1 + n_2^* \cos\theta_2 + \frac{4\pi}{c}\hat{\sigma}_2)^2}$$

It is seen, that in the case of TE-polarization the law of energy conservation takes place:

$$R_{(s)} + T_{(s)} + I_{(s)}^{(j)} = 1$$

Write the analogical expressions for TM-polarization:

$$J_{(p)}^{(i)} = \frac{cn_1}{4\pi}|E_{ip}|^2 \cos\theta_1$$

$$J_{(p)}^{(R)} = \frac{cn_1}{4\pi}|E_{ip}|^2 \cos\theta_1 \cdot \frac{((n_2^* + \frac{4\pi}{c}\hat{\sigma}_2)\cos\theta_1 - n_1 \cos\theta_2)^2}{((n_2^* + \frac{4\pi}{c}\hat{\sigma}_2)\cos\theta_1 + n_1 \cos\theta_2)^2}$$

$$J_{(p)}^{(t)} = \frac{c(n_2^* + i\kappa_2^*)}{4\pi}|E_{ip}|^2 \cos\theta_2 \cdot \frac{4n_1^2 \cos^2\theta_1}{((n_2^* + \frac{4\pi}{c}\hat{\sigma}_2)\cos\theta_1 + n_1 \cos\theta_2)^2}$$

$$I_{(p)}^{(j)} = (\hat{\sigma}_2 - i\frac{c}{4\pi}\kappa_2^*) \cdot |E_{ip}|^2 \cdot \cos\theta_2 \cdot \frac{4n_1^2 \cos^2\theta_1}{((n_2^* + \frac{4\pi}{c}\hat{\sigma}_2)\cos\theta_1 + n_1 \cos\theta_2)^2}$$

The energy distribution:

$$R_{(p)} = \frac{((n_2^* + \frac{4\pi}{c}\hat{\sigma}_2)\cos\theta_1 - n_1 \cos\theta_2)^2}{((n_2^* + \frac{4\pi}{c}\hat{\sigma}_2)\cos\theta_1 + n_1 \cos\theta_2)^2}$$

$$T_{(p)} = \frac{4n_1 \cdot (n_2^* + i\kappa_2^*)\cos\theta_1 \cos\theta_2}{((n_2^* + \frac{4\pi}{c}\hat{\sigma}_2)\cos\theta_1 + n_1 \cos\theta_2)^2}$$

$$I_{(p)}^{(j)} = \cdot \frac{\frac{16\pi}{c} n_1 (\hat{\sigma}_2 - i\frac{c}{4\pi}\kappa_2^*)\cos\theta_1 \cos\theta_2}{((n_2^* + \frac{4\pi}{c}\hat{\sigma}_2)\cos\theta_1 + n_1 \cos\theta_2)^2}$$

It is seen, that in the case of TM-polarization, the law of energy conservation takes place as well:

$$R_{(p)} + T_{(p)} + I_{(p)}^{(j)} = 1$$

# REFERENCES

1. T.A.Kudykina, M.P.Lisitsa. Optoelectronics and Semiconductor Technics,(russ.) v.**32**, 106, 1997. T.A.Kudykina. SPIE, **3573,** 324, 1998.

2. M.P.Lisitsa, N.G.Tsvelyh, Optika i spektroskopia. **2**(5), 674-676 (1957); **4**(3), 373-377 (1958); **5**(5), 622-624 (1958); **7**(4), 552-557 ( 1959).

   M.P.Lisitsa, V.M.Maevskii, N.G.Tsvelyh. Opt.spektr. **5** (2), 179-183 (1958).

3. A.H.Pfund. JOSA, **23**, 270, 1933.

4. R.Schulze. Phys.Zs., **34**, 24, 1933.

5. L.T.Canham, Appl.Phys.Lett.**57**, 1046, 1990.

6. V.Lehmann, U.Gösele, Appl.Phys.Lett.**58**, 856, 1991.

7. P.P.Ewald. Rev.Mod.Phys. **37**(1), 46, 1965.

8. C.W.Oseen. AnnPhys.**48**,1,1915.

9. M.Born, E.Wolf. Principles of Optics. Pergamon Press, Oxford-Frankfurt, 1968, 719 p.

10. V.Weisskopf."How light interact with matter" , in Lasers and Light, (W.H.Freeman, San Francisco, 1969).

11. H.Fearn, D.F.V.James, P.W.Milonni. Am.J.Phys. **64**, 986, 1996.

12. R.C.Faust. Phil.Mag. **41**, 1238, 1950.

13. H.Murmann. Zs.f.Phys. **30**, 161, 1933.

14. P.Rouard, P.Bousquet, , Progress in Optics. Ed..Wolf E., **4**, p.145-197 (1965).

15. G.Hass, J.E.Waylonis. JOSA, **51**, 719-722, 1961.

16. J.A.Stratton. Electromagnetic Theory. N.Y., 1941.

17. H.Mayer, J.de Physic.**25** (1-2), 172-182 , 1964.

18. J.-J.Xu, J-F.Tang, Appl.Opt. **28** (14), 2925-2928 (1989).

19. M.Yamamoto,T.Namioka, Appl.Opt.**31** (10), 1612-1621 (1992).

20. T.A.Kudykina, phys. stat. solidi.**(b)160**, 365-373 (1990);

21. T.A.Kudykina, phys. stat. solidi **(b) 165** (2), 591-598 (1991).

22. В.Л.Винецкий, Т.А.Кудыкина. ФТП, **23**, 1910-1913, 1089 (V.L.Vinetskii, T.A.Kudykina. Physics and Technology of Semiconductors, russ.).

23. N.Fang, H.Lee, Ch.Sun, X.Zhang. Science. **308**, 524, 2005.

24. A.J.Hoffman, L.Alekseyev, S.Howard, K.J.Franz, D.Wasserman, V.A.Podolskiy, E,E,Gmachl. Nature Materials. **6**, 946, 2007.